\documentclass[superscriptaddress,showpacs,aps,twocolumn,prb,floatfix]{revtex4-1}
\usepackage{amssymb}
\usepackage{amsmath}
\usepackage[dvips]{graphicx}
\usepackage{bm}

\newcommand{\bea}{\begin{eqnarray}}
\newcommand{\eea}{\end{eqnarray}}

\def\beq{\begin{equation}}
\def\eeq{\end{equation}}

\begin{document}

\title{Thermopower of an SU(4) Kondo resonance under an SU(2) symmetry-breaking field}

\author{P. Roura-Bas}
\affiliation{Dpto de F\'{\i}sica, Centro At\'{o}mico Constituyentes, Comisi\'{o}n
Nacional de Energ\'{\i}a At\'{o}mica, Buenos Aires, Argentina}
\author{L. Tosi}
\affiliation{Centro At\'omico Bariloche and Instituto Balseiro, Comisi\'on Nacional de
Energ\'{\i}a At\'omica, 8400 Bariloche, Argentina}
\author{A. A. Aligia}
\affiliation{Centro At\'omico Bariloche and Instituto Balseiro, Comisi\'on Nacional de
Energ\'{\i}a At\'omica, 8400 Bariloche, Argentina}
\author{P. S. Cornaglia}
\affiliation{Centro At\'omico Bariloche and Instituto Balseiro, Comisi\'on Nacional de
Energ\'{\i}a At\'omica, 8400 Bariloche, Argentina}

\date{\today }

\begin{abstract}
We calculate the thermopower of a quantum dot  described by two doublets 
hybridized with two degenerate bands of two conducting leads, conserving orbital (band) and spin
quantum numbers,  
as a function of the temperature $T$ and a splitting $\delta$ of the quantum dot levels which breaks the SU(4) symmetry. The splitting  can be regarded
as a Zeeman (spin) or valley (orbital) splitting.
We use the non-crossing approximation (NCA), the slave bosons in the mean-field approximation (SBMFA)
and also the numerical renormalization group (NRG) for large $\delta$.
The model describes transport through clean C nanotubes 
and in Si fin-type field effect transistors, under an applied magnetic field. 
The thermopower as a function of temperature $S(T)$ displays two dips that
correspond to the energy scales given by the Kondo temperature $T_K$ and  
$\delta$ and one peak when $k_BT$ reaches the charge-transfer energy. 
These features are much more pronounced 
than the corresponding ones in the conductance, indicating that the thermopower
is a more sensitive probe of the electronic structure at intermediate or high energies.
At low temperatures ($T \ll T_K$) $T_K S(T)/T$ is a constant that increases strongly near the degeneracy point $\delta=0$. 
We find that the SBMFA fails to provide an accurate description of the thermopower for large $\delta$. 
Instead, a combination of Fermi liquid relations with the quantum-dot occupations calculated within the NCA gives
reliable results for $T \ll T_K$.

\end{abstract}

\pacs{72.20.Pa, 72.15.Qm,73.23.Hk, 73.63.Nm}

\maketitle


\section{Introduction}

\label{intro}

Materials with potentially useful thermoelectric properties and in particular large thermopower,
are currently a subject of intense research
due to their potential applications, for example in refrigeration or
conversion of waste heat into electricity.\cite{tritt,sales,tera,bent,kana,klie,maso}
In addition, the thermopower is a useful tool to obtain
additional insight on fundamental problems, like the Kondo 
effect.\cite{bic,velmo,cost,vel}
The focus of research on thermal and thermoelectric properties has moved in the last years to 
nanostructures such as quantum dots,\cite{vel,harm,sche,boese,dong,kim}
carbon nanotubes,\cite{hone,yu} molecules \cite{red} nanowires, 
and spin systems.\cite{sogo,sogo2,roz,arra}
As we will show, the thermopower can be a more useful tool than the conductance,
to study some features of the electronic structure of Kondo systems at finite energies. 

Electronic and thermal transport through single level quantum dots is well described by the SU(2)
Anderson model, and
the Seebeck coefficient
(thermopower) has been calculated using this model.\cite{vel,boese,dong,kim}
In particular, Costi and Zlati\'{c} have made a comprehensive study of
the transport properties using the numerical renormalization group (NRG).\cite{vel} 
A particularly interesting multilevel system is
the SU(4) Anderson model, which describes quantum dots in 
carbon nanotubes,\cite{lim,ander,lipi,buss,fcm,grove} and silicon nanowires.\cite{tetta} In
contrast to the SU(2) case, the electron spectral density near the Fermi
level of the SU(4) Anderson model in the strong coupling (Kondo) limit for
nearly one electron at the dot is highly asymmetric,\cite{lim,fcm} leading
to a high derivative at the Fermi energy, which (neglecting phonon-drag effects) in turn is proportional to
the Seebeck coefficient $S$ at low temperatures.\cite{vel} 

The interest in Si
nanowires increased due to the possibility of reducing the thermal
conductance $\kappa $, leading to a large figure of merit $\sigma
S^{2}T/\kappa $, where $\sigma $ is the conductivity.\cite{he} Recently, the
conductance of Si fin-type field effect transistors has been measured under
an applied magnetic field $B$, which leads to a 
crossover from an SU(4) to an
orbital SU(2) Kondo effect.\cite{tetta} The results were interpreted using
the non-crossing approximation (NCA). In C nanotubes, even for $B=0$, there might be a symmetry reduction
to SU(2) due to disorder-induced intervalley mixing.\cite{grove}

In this work, we calculate the thermopower of the Anderson impurity model
for two doublets, each one either spin or orbital degenerate, 
and infinite on-site Coulomb repulsion in the
Kondo limit, as a function of the level splitting $\delta$, which 
corresponds to the Zeeman splitting for two orbitally degenerate levels in presence of
an applied magnetic field, as in Si fin-type field effect transistors.\cite{tetta} 
The symmetry of the model is SU(4) for $\delta =0$ and SU(2) for $\delta \neq 0$. 
We use complementary theoretical approaches: the NCA,\cite{bic,velmo,fcm}
slave bosons in the mean-field approximation (SBMFA),\cite{col,hew,lady,soc} and
Fermi liquid theory for low temperatures. We also use NRG to test the other
approaches in the SU(2) limit. 

The SBMFA satisfies Fermi liquid
relationships at $T=0$ and is expected to capture the low-energy physics in
the Kondo limit. In turn, while NCA has an error of about 15\% in the value
of the spectral density at very low temperature according to the Friedel sum
rule,\cite{fcm} it describes better the whole behavior, as shown for example
in a previous comparison of NCA and NRG results in the SU(2) case.\cite{compa} 
In addition, if the differential conductance $G$ is normalized at 
$T=0$, the leading behavior of  $G$ for small voltage and temperature 
\cite{roura} agrees with alternative Fermi liquid 
approaches,\cite{ogu,rpt,sela,scali} and the temperature dependence of the conductance
practically coincides with the NRG result over several decades of
temperature.\cite{roura} 
Calculations of the thermopower for more than one spin-degenerate level within the
NCA, compared well with experiments in some Ce
compounds.\cite{bic,velmo} 
NRG is a very accurate technique at low temperatures.\cite{bulla} However, 
for two bands, the Hilbert
space is increased 16 times in each iteration, instead of 4 times for one band, 
making the technique much more demanding if the same degree of
accuracy is wished. 
In addition, the SU(4) symmetry cannot be used to reduce the size of the matrices, since
this symmetry is broken by $\delta$. 
Therefore we use here NRG only in the limit 
$\delta \rightarrow +\infty$, in which only one doublet and the band that mixes with it
play a significant role.

The paper is organized as follows. In Section \ref{model} we
describe the model used. The approximations and the equation for the Seebeck coefficient
are described in Section \ref{forma}. The numerical results are presented in Section \ref{res}. 
Section \ref{sum} contains a summary and a short discussion. 

\section{Model}

\label{model}

We start with a generalization of the Anderson model for infinite on-site
Coulomb repulsion, which contains a singlet $|s\rangle $ with $\mathcal{N}$ (even)
particles and two spin doublets $|i\sigma \rangle $ ($i=1,2$ is the valley
index; $\sigma =\uparrow $ or $\downarrow $) with $\mathcal{N}+1$ (or $\mathcal{N}-1$) particles
representing the four spin and valley degenerate states of a quantum dot created for
example by depleting the density at two points of a C nanotube or in a Si
nanowire. The dot is connected to two conducting leads which are also spin and valley degenerate. 
The SU(4) symmetry
is then broken by applying a magnetic field or breaking the valley
degeneracy in a simple way. In both cases, interchanging spin ($\sigma$) and valley ($i$)
indices if necessary, the Hamiltonian can be written in the form 
\begin{eqnarray}
H &=&E_{s}|s\rangle \langle s|+\sum_{i\sigma }E_{i}|i\sigma \rangle \langle
i\sigma |+\sum_{\nu k\sigma }\epsilon _{\nu k}c_{\nu k i \sigma }^{\dagger
}c_{\nu k i \sigma }  \notag \\
&&+\sum_{i\nu k\sigma }(V_{\nu}|i\sigma \rangle \langle s|c_{\nu k i \sigma }+\mathrm{H.c}.),  \label{ham}
\end{eqnarray}%
where $c_{\nu k i \sigma }^{\dagger }$ create conduction states at the left 
($\nu =L$) or right ($\nu =R$) lead, and $V_{\nu}$ is the hopping between the lead $\nu$
and both doublets, assumed independent of $k$ for simplicity. 
The doublets are split by an energy $\delta=E_{2}-E_{1}$. This corresponds to the Zeeman 
splitting when the SU(4) symmetry of the model for $\delta=0$ is broken by an applied magnetic field.

While there are four spin degenerate bands of mobile electrons, 
depending on valley index $i$
or position with respect to the quantum dot (left or right), 
for each energy $\epsilon _{L k}= \epsilon _{R k'}$ for which 
there are states at the left and the right, only the linear combination 
$V_{L k} c_{L k i \sigma }^{\dagger } + V_{R k'} c_{R k' i \sigma }^{\dagger }$ 
hybridizes with the state $|i\sigma \rangle $. Thus, the model is effectively a two-band model.

We note that in the case of intervalley mixing induced by disorder in C nanotubes, the
hopping elements for small magnetic field depend on the valley and lead indices.\cite{grove} 
In this case, the formalism used in this work
is not applicable. It seems that a full non-equilibrium formalism is needed
to treat the most general case, as that developed for the conductance in
Ref. \onlinecite{benzene} and sketched in Ref. \onlinecite{desint}. The model is not
applicable either for the case of magnetic impurities in C nanotubes, for
which symmetry-breaking geometrical effects play an important role.\cite{baru}

\section{The formalism}

\label{forma}

\subsection{Equations for the thermopower}

\label{thermo}

In the limit of vanishing applied bias voltage and temperature difference between the leads, 
the electronic part of the transport coefficients can be
evaluated in terms of the total spectral density at the dot 
$\rho_d (\omega)=\sum\limits_{i\sigma }\rho _{i\sigma }(\omega )$. 
This is possible due to the fact that the couplings between the quantum dot 
and the right or left leads are proportional.\cite{meir}
If this were not the case (as in nanotubes affected by disorder 
\cite{grove} or for some molecules \cite{benzene}) a different
formalism would be needed.\cite{benzene,desint}

The Seebeck coefficient is simply given by \cite{vel}

\begin{equation}
S=\frac{-I_{1}(T)}{eTI_{0}(T)},  \label{see}
\end{equation}%
where $e$ is the absolute value of the electronic charge and

\begin{equation}
I_{n}=\int \omega ^{n}\rho_d (\omega )\left( -\frac{\partial f}{\partial
\omega }\right) d\omega ,  \label{in}
\end{equation}%
where $f(\omega )$ is the Fermi function, and the zero of energy
is taken at the Fermi energy $\epsilon_F=0$

\subsection{Approximations for the spectral density}

\label{apro}

To calculate the total spectral density $\rho_d (\omega )$ that enters 
Eqs. (\ref{in}) we have used mainly the NCA. At $T=0$ have used the SBMFA and also
a combination of Fermi liquid relationships and quantum-dot occupations obtained using NCA.
In the limit $\delta \rightarrow + \infty $, we have also used NRG to 
shed light on the virtues and shortcomings of the other approximations.

In the NCA and SBMFA, an auxiliary boson $b$, and four auxiliary
fermions $f_{i\sigma }$ are introduced, so that the localized states are
represented as
\begin{equation}
|s\rangle =b^{\dag }|0\rangle \text{, }|i\sigma \rangle =f_{i\sigma }^{\dag
}|0\rangle ,  \label{rep}
\end{equation}%
where $|0\rangle $ is the vacuum. These pseudoparticles should satisfy the
constraint 
\begin{equation}
b^{\dag }b+\sum_{i\sigma }f_{i\sigma }^{\dag }f_{i\sigma }=1.  \label{cons}
\end{equation}%
The NCA solves a system of self-consistent equations to obtain the Green
functions of the auxiliary particles, which is equivalent to
sum an infinite series of diagrams (all those without crossings) in the
corresponding perturbation series in the hopping, and afterwards a
projection on the physical subspace of the constraint is made. The formalism of
the NCA for this problem or similar ones is described in previous 
papers.\cite{bic,velmo,lim,benzene} In particular, the more general case of complex
hoppings is treated in Ref. \onlinecite{benzene}. Therefore, we do not give more details
here. The application of NRG in the case of only one doublet ($\delta
\rightarrow +\infty $) has also been explained before.\cite{vel}

In the SBMFA, the boson operators are replaced by a number $b_{0}=\langle
b\rangle $, and the energy is minimized with respect to $b_{0}$ and a
Lagrange multiplier $\lambda $ that enforces Eq. (\ref{cons}). Assuming a
constant density of unperturbed conduction electrons $\rho$ extending
from $-D$ to $D$ and filled to the Fermi level $\epsilon _{F}=0$, a simple
generalization of the case of one doublet,\cite{hew,lady} leads to the
following change of energy after introduction of the impurity

\begin{eqnarray}
E &=&\frac{1}{\pi } \sum\limits_{i}\left[ \tilde{\Delta}\ln \left( \frac{%
\tilde{E}_{i}^{2}+\tilde{\Delta}^{2}}{D^{2}}\right) +2\tilde{E}_{i}\arctan
\left( \frac{\tilde{\Delta}}{\tilde{E}_{i}}\right) \right]  \notag \\
&&-\frac{4\tilde{\Delta}}{\pi }+\lambda (b_{0}^{2}-1),  \label{esb}
\end{eqnarray}%
where $\tilde{E}_{i}=E_{i}+\lambda $, $\tilde{\Delta}=b_{0}^{2}\Delta $, and 
$\Delta = \pi \rho (V_L^{2}+V_R^{2})$ is the total half resonant level width (adding the
contributions from both leads). Above $D\gg \tilde{\Delta}$ was
assumed. Minimizing $E$ with respect to $\lambda $ one obtains an equation
that allows to relate $\tilde{E}_{1}$ (or $\lambda $) with half the
quasiparticle level width $\tilde{\Delta}$ (which is of the order of the
Kondo temperature). After some algebra we obtain

\begin{eqnarray}
\tilde{E}_{1} &=&\left[ \frac{\delta ^{2}}{4}+(1+\beta ^{2})\tilde{\Delta}%
^{2}\right] ^{1/2}+\beta \tilde{\Delta}-\frac{\delta }{2},  \notag \\
\beta ^{-1} &=&\tan \left[ \frac{\pi }{2}\left( 1-\frac{\tilde{\Delta}}{%
\Delta }\right) \right] .  \label{e1}
\end{eqnarray}%
Minimizing $E$ with respect to $\tilde{\Delta}$ one obtains

\begin{equation}
\frac{1}{\pi }\ln \frac{(\tilde{E}_{1}^{2}+\tilde{\Delta}^{2})((\tilde{E}%
_{1}+\delta )^{2}+\tilde{\Delta}^{2})}{D^{4}}+\frac{\tilde{E}_{1}-E_{1}}{%
\Delta }=0,  \label{delt}
\end{equation}
and replacing $\tilde{E}_{1}$ from Eq. (\ref{e1}) in Eq. (\ref{delt}) an
equation for the single unknown $\tilde{\Delta}$ is obtained, which we solve
numerically.

The occupation and the spectral density near the Fermi energy for each
doublet are

\begin{eqnarray}
n_{i\sigma } &=&\langle |i\sigma \rangle \langle i\sigma |\rangle =\frac{1}{%
\pi }\arctan \left( \frac{\tilde{\Delta}}{\tilde{E}_{i}}\right) ,  \notag \\
\rho _{i\sigma }(\omega ) &=&\frac{b_{0}^{2}\tilde{\Delta}/\pi }{(\omega -%
\tilde{E}_{i})^{2}+\tilde{\Delta}^{2}}.  \label{nrho}
\end{eqnarray}

\subsection{Fermi liquid theory}

\label{fermi}

\bigskip For an SU(N) model which channel index $j=1$ to $N$, and a simple
symmetry breaking perturbation (like a generalized magnetic field) that does
not mix channels, so that spin and channel are conserved, the Friedel sum rule relates 
the spectral density at the
Fermi level $\epsilon _{F}$ with the number of displaced electrons for each
channel.\cite{yoshi2}
The latter coincides with the occupation for a constant
unperturbed density of conduction electrons with a wide band $D\gg \tilde{\Delta}$ as we assume,
where $\tilde{\Delta}$ (of the order of the Kondo temperature $T_{K}$) 
is the resonant level width of the quasiparticles [as in the SBMFA, Eq. (\ref{nrho})] 

Then one has \cite{fcm,yoshi2}

\begin{equation}
\rho _{j}(\epsilon _{F})=\frac{1}{\pi \Delta }\sin ^{2}(\pi n_{j}).
\label{fl1}
\end{equation}%
In addition, the derivatives at $\epsilon _{F}$ are also known (using for
example renormalized perturbation theory) \cite{scali,yoshi,kirchner}, 

\begin{equation}
\frac{\partial \rho _{j}(\omega )}{\partial \omega }|_{\epsilon _{F}}=\sin
(2\pi n_{j})\frac{\rho _{j}(\epsilon _{F})}{\tilde{\Delta}}.  \label{fl2}
\end{equation}

From Eqs. (\ref{nrho}), it is apparent that the SBMFA (in which we have
chosen $\epsilon _{F}=0$) satisfies these relationships.

Using Eqs. (\ref{fl1}) and (\ref{fl2}) and a Sommerfeld expansion in Eqs (%
\ref{in}), the Seebeck coefficient Eq. (\ref{see}) for $T\rightarrow 0$
becomes

\begin{equation}
S=-\frac{2\pi ^{2}k_{B}T}{3\tilde{\Delta}}\frac{\sum_{i\sigma }\sin ^{3}(\pi
n_{i\sigma })\cos (\pi n_{i\sigma })}{\sum_{i\sigma }\sin ^{2}(\pi
n_{i\sigma })}.  \label{sfl}
\end{equation}

\section{Numerical results}

\label{res}

Without loss of generality, we take $\epsilon _{F}=E_{s}=0$, where $\epsilon _{F}$ is the
Fermi level of the leads.
For the numerical calculations, we assume a constant density of states per
spin of the leads $\rho= 1/(2D) $ between $-D$ and $D$. We take the unit of energy
as the total level width of both doublets $\Gamma =2 \Delta =2 \pi \rho (V_L^{2}+V_R^{2})$. 
The energy of both doublets is denoted as $E_1 = E_d$, $E_2 = E_d + \delta$. 

\subsection{The limit of one doublet}

\label{su2}

For $\delta \rightarrow +\infty $, the model is equivalent to the limit of 
infinite on-site repulsion $U$  
of the simplest Anderson model, studied in detail before using the
NRG.\cite{vel} Here we compare results obtained with NRG with those of NCA
and SBMFA to see the limitations of the different methods, which will be
useful for the analysis of the general case. 

\begin{figure}[tbp]
\includegraphics[width=8.0cm]{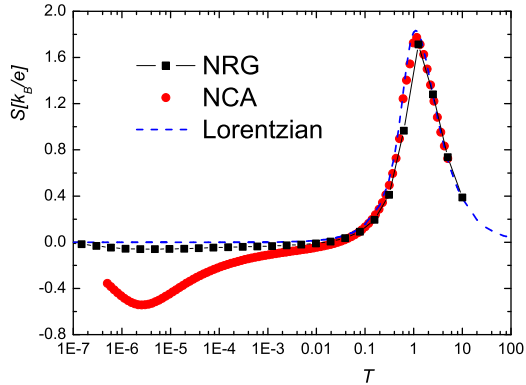} 
\caption{(Color online) Thermopower as a function of temperature for
$E_d=-4 \Gamma$ and $\delta \rightarrow +\infty$. Squares: NRG, dots: NCA,
dashed line: Lorentzian spectral density (see text).}
\label{fsu2}
\end{figure}

In Fig. \ref{fsu2} we display the thermopower as a function of temperature in
the Kondo regime. There is a dip near the Kondo temperature $T_K$ due to the fact that 
the Kondo peak in the spectral density is slightly above the Fermi level $\epsilon_F$, and a peak near the 
charge-transfer energy $\epsilon_F -E_d$ due to the corresponding peak below $\epsilon_F$.
We mean ``near'' as an estimation of the order of magnitude. For example, the maximum of the curve
within NCA is at $T=1.10 \Gamma$, while $\epsilon_F -E_d= 4 \Gamma$ in the figure.

The spectral density $\rho_d(\omega)$ of the Anderson model for only one doublet
($\delta \rightarrow +\infty$) 
is well known. 
In particular, using a local-moment approach,
Logan and coworkers have shown that in the Kondo regime, the charge-transfer peaks 
(which carry most of the spectral weight in this regime) have a Lorentzian shape of a width two times 
larger than that of the non-interacting case.\cite{logan} We have included in Fig. \ref{fsu2}
the thermopower that results replacing $\rho_d(\omega)$ in Eqs. (\ref{see}) and (\ref{in}), 
by a simple Lorentzian of total width $2 \Gamma$ centered at $E_d$. The agreement with the NCA result 
at high temperatures is remarkable. 
Instead, the charge-transfer peak in the thermopower within NRG has a small shift to higher temperatures 
and an intensity about 10 \% lower. We have verified that this also happens for other values of $E_d$.
This is probably related with resolution problems of the NRG at large energies, as discussed below.  

At low temperatures, the NRG results are more reliable than those of NCA and SBMFA. We have extracted 
the linear term in $T$ in the Seebeck coefficient, plotting $S/T$ vs $T$ for small $T$ and looking
at the extrapolation to $T=0$. The results are shown in Table I. They have an uncertainty of the order of 10 \%. 
We have also calculated the conductance
$G(T)$ (not shown) and the total occupancy $n$ adding both spins. 
Using the results for $G(T)$, we estimated the Kondo temperature from the requirement that $G(T_{K}^{G})=G(0)/2$.
Using $n_{1 \uparrow}=n_{1 \downarrow}=n/2$, $n_{2 \uparrow}=n_{2 \downarrow}=0$, and 
the Fermi liquid expression Eq. (\ref{sfl}),
$S/T$ is calculated in an independent way. Taking into account the exponential variation of $T_K$ with
$E_d$ one can see a semiquantitative agreement between both results in Table 1. The remaining quantitative
discrepancy can be ascribed to the difference between $T_{K}^{G}$ and $\tilde{\Delta}$ as a measure of
the Kondo temperature. In fact from renormalized perturbation theory one obtains $T_{K}^{G} / \tilde{\Delta}=0.746$
in the extreme Kondo limit while this ratio increases beyond 1 when valence fluctuations are allowed,\cite{rpt} 
and the Wilson ratio decreases.\cite{ogu,rpt,sela,scali}

\begin{table}[t]
\caption{Kondo temperature from $G(T)$ ($T_{K}^{G}$), total occupation of the lowest doublet ($n$), 
linear coefficient of $S(T)$ from NRG ($ST_{K}^{G}/T$) and from a Fermi liquid theory ($S\tilde{\Delta}/T$)
for different values of the charge-transfer energy $\epsilon_F -E_d$.}

\begin{tabular}{||l||l|l|l|l||}
\hline\hline
$E_{d}/\Gamma $ & $T_{K}^{G}/\Gamma $ & $n$ & $ST_{K}^{G}/T$ & $S\tilde{\Delta}/T$ \\ \hline\hline
-1 & $2.22\times 10^{-2}$ & 0.732 & -3.48 & -2.450 \\ \hline
-2 & $6.15\times 10^{-4}$ & 0.897 & -0.93 & -1.047 \\ \hline
-3 & $2.22\times 10^{-5}$ & 0.941 & -0.51 & -0.611 \\ \hline
-4 & $8.38\times 10^{-7}$ & 0.958 & -0.38 & -0.437 \\ \hline
-5 & $3.16\times 10^{-8}$ & 0.967 & -0.27 & -0.339 \\ \hline
-6 & $1.18\times 10^{-9}$ & 0.973 & -0.22 & -0.317 \\ \hline\hline

\end{tabular}
\end{table}

The NCA has the drawback that it does not fulfill Fermi liquid relationships.
For example, the spectral density at the Fermi energy $\rho_d(0)$ at temperatures well below $T_K$ differs by about 
10 or 20\% in the Kondo regime from the value predicted by the Friedel sum rule.\cite{fcm}
A detailed comparison of $\rho_d(\omega)$ between NRG and NCA in the Kondo regime is given by 
Fig. 10 of Ref. \onlinecite{compa}. One can see that in addition to the larger value of
$\rho_d(0)$, the spectral density is more asymmetric for the NCA. This is probably the main reason 
of the factor near five between the magnitude of the dip in $S(T)$ near $T_K$ calculated with NCA 
with respect to the NRG result. Part of the discrepancy is probably also due to lack of resolution at finite energies
within the NRG. For example in models of two quantum dots, the split Kondo peaks
in the spectral density are considerably broadened, losing intensity (Fig. 11 of Ref. \onlinecite{vau}).
Another example is the plateau in the conductance $G(T)$ 
observed at intermediate temperatures $T$ in transport through C$_{60}$ molecules
for gate voltages for which triplet states are important,\cite{roch,serge}
which was missed in early NRG studies, but captured by the NCA.\cite{st1,st2}
More recent NRG\ calculations using tricks to improve the resolution,\cite{freyn} 
have confirmed this plateau.\cite{serge} 
 
In any case, the one-level SU(2) limit is the worst case for the NCA,
because the real spectral density tends to be symmetric, while the NCA improves with
increasing N for SU(N) models.\cite{bic} 
 
\begin{figure}[tbp]
\includegraphics[width=7.5cm]{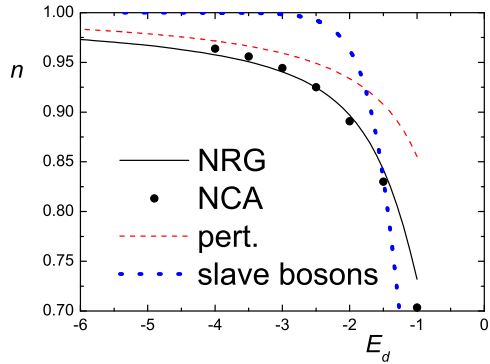} 
\caption{(Color online) Total occupancy as a function of $E_d$ for $\delta \rightarrow +\infty$ 
and different approximations.}
\label{ocup}
\end{figure}

The absolute value of $S$ is exaggerated within NCA for $T \rightarrow 0$.
However, as shown in Fig. \ref{ocup}, there is a good agreement between the occupancies 
calculated with NRG and NCA. This suggests that using the occupancies calculated with
NCA and Eq. (\ref{sfl}) a semiquantitatively correct result for the linear
part of $S(T)$ as $T \rightarrow 0$ can be obtained. Instead, while the
SBMFA satisfies Fermi liquid relationships, the occupancies are not accurate for 
$\delta \rightarrow +\infty$. Even perturbation theory up to second order 
in $V_{\nu}$ neglecting spin flip leads to a better result for $n$, although
(in contrast to SBMFA) this approach is unable to predict the right magnitude of $T_K$.

\subsection{Temperature dependence in the general case}

\label{temp}

\begin{figure}[tbp]
\includegraphics[width=8.0cm]{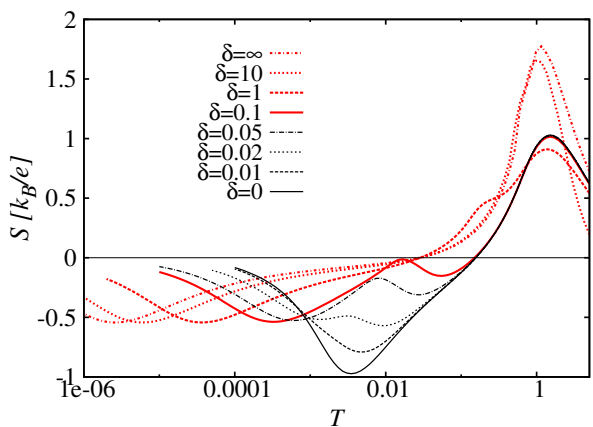} 
\caption{(Color online) Thermopower as a function of temperature for $E_d=-4 \Gamma$ and different values 
of the level splitting.}
\label{seet}
\end{figure}

In Fig. \ref{seet} we represent the NCA results for the Seebeck coefficient $S(T)$ as a function of temperature for different values
of $\delta$, which represents the splitting due to a Zeeman term for example. The spectral density $\rho_d(\omega)$
for finite $\delta$ has been studied before.\cite{lim,fcm}
At $\delta=0$, $\rho_d(\omega)$ has a peak just above the Fermi energy $\epsilon_F$ with width of the order of
$2T^4_K$, where $T^4_K$ is the Kondo temperature of the SU(4) limit $\delta = 0$.
For the parameters of Fig. \ref{seet},  $T^4_K$ is of the order of $0.01 \Gamma$.\cite{fcm}
As $\delta$ increases above $T^4_K$, the Kondo peak splits and another peak at an energy of the order
of $\delta$ above $\epsilon_F$ appears.
This peak above $\epsilon_F$ originates a dip
in $S(T)$ for $T \sim \delta$.
Note that peaks in $\rho_d(\omega)$ above $\epsilon_F$ lead to a positive contribution to $I_1$ [see Eq. (\ref{in})],
which in turn lead to a negative contribution to $S(T)$ at the corresponding temperature [see Eq. (\ref{see})]
Most of the spectral weight of the spectral density lies in the charge-transfer peak at energy $E_d$, which lies below
the Fermi energy. Thus, when the temperature reaches values of the order of the charge-transfer energy $\epsilon_F -E_d$, 
the thermopower becomes positive [$I_0 >0$, $I_1 <0$ in Eq. (\ref{see})]. 

As discussed above, for $\delta \rightarrow +\infty$ (the case of one SU(2) doublet), $S(T)$ shows one dip
at $T_K \sim 10^{-6} \Gamma$ and a peak near the charge-transfer energy. For 
$\delta =0$, these qualitative features remain, but the Kondo temperature is four orders of magnitude
larger, and the dip near $T^4_K$ is more pronounced, due to the larger asymmetry of the peak in the spectral density
with respect to $\epsilon_F$,\cite{fcm} leading to a larger integral $I_1$ [see Eqs. (\ref{see}) and (\ref{in})].
Since $T_K$ changes with $\delta$, as discussed below, the dip at $T_K$ displaces towards larger temperatures as $\delta$ decreases.
Instead, the charge-transfer peak remains approximately at the same temperature and decreases in magnitude due mainly to additional 
broadening of the corresponding peak in the spectral density $\rho(\omega)$ as the SU(4) limit is approached, and also due to some transfer 
of the spectral weight to the Kondo peak. 
While this transfer is not large, the Kondo peak has a larger 
weight in the integrals $I_n$ due to the factor of the derivative of the Fermi function [see Eqs. (\ref{see}) and (\ref{in})].

For $\delta > T^4_K$, for example $\delta = 0.02 \Gamma$ in Fig. \ref{seet}, an additional dip develops due 
to the peak near $\delta$ in $\rho_d(\omega)$. As $\delta$ increases further, the dip moves to higher temperatures, 
as it is apparent for $\delta=0.05$ and 0.1 in the figure. The relative minimum of $S(T)$ near the dip lies at temperatures of the order of $\delta$
but smaller, probably because of the large intensity of the peak near the charge-transfer energy, which pushes this minimum to lower
temperatures. For $\delta=1$ this dip turns to a shoulder at the left of the charge-transfer peak and for larger $\delta$, the dip 
and the peak cross, interchanging the order of temperatures for which they appear.

\begin{figure}[tbp]
\includegraphics[width=8.0cm]{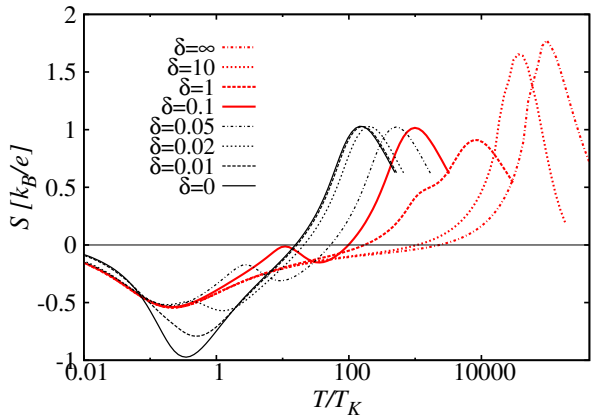} 
\caption{(Color online) Thermopower as a function of $T/T_K$ for $E_d=-4 \Gamma$ and different values 
of the level splitting.}
\label{sttk}
\end{figure}

In Fig. \ref{sttk} we show the thermopower as a function of $T/T_K(\delta)$, where
$T_K(\delta)$ is the Kondo temperature for each value of $\delta$. Here we determined $T_K(\delta)$ from the
requirement that the conductance $G(T_K(\delta))=e^2/h$ for symmetric leads ($V_L=V_R$).  
To a good degree of accuracy, it is given 
by the expression
\begin{equation}
T_K(\delta)=\left\{ (D+\delta )D\exp \left[ \frac{\pi E_{1}}{2\Delta}\right] + 
\frac{\delta^2}{4} \right\} ^{1/2}-\frac{\delta}{2},  \label{tk}
\end{equation}
obtained from a simple variational wave function.\cite{desint,fcm}
With increasing $\delta$,
$T_{K}$ stays roughly constant until $\delta > T^4_K = T_K(0)$ and then it decreases strongly.
From Fig. \ref{sttk} it is apparent that the dip at smaller temperatures remains
at $T \sim T_K(\delta)$ for all values of $\delta$. 
One can also see that the magnitude of the dip
and (the absolute value of the thermopower for $T \sim T_K$) increases as $\delta$ decreases, being maximum
at the SU(4) point $\delta=0$. Note that while the magnitude of this dip is exaggerated by the NCA 
for $\delta \rightarrow +\infty$, as explained in the previous section, we believe that this is not the case 
for $\delta=0$, because the spectral density is naturally asymmetric in this case, and since the NCA
is a 1/N expansion, its accuracy improves with N in SU(N) models.\cite{bic} Furthermore NCA calculations
of the thermopower of Ce compounds, in which orbital degeneracy is important, compared well with experiment.\cite{bic,velmo} 

\begin{figure}[tbp]
\includegraphics[width=8.0cm]{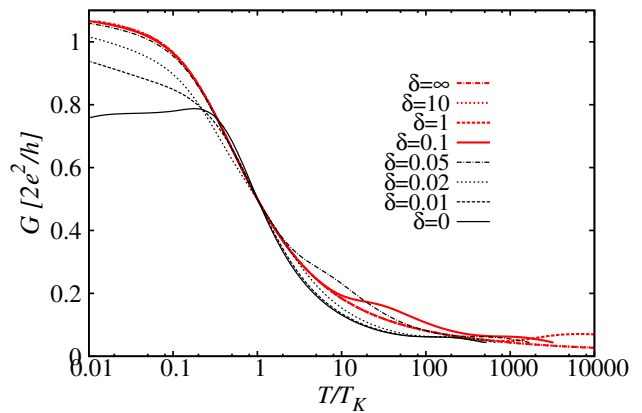} 
\caption{(Color online) Conductance  as a function of $T/T_K$ for $E_d=-4 \Gamma$ and different values 
of the level splitting.}
\label{g}
\end{figure}

In Fig. \ref{g} we show the conductance $G(T)$  calculated with the NCA for different values of $\delta$.
$G(0)$ is slightly larger for $\delta \rightarrow +\infty$ and lower for $\delta = 0$ with respect to
the correct values due to the failure of the NCA spectral density to satisfy Friedel sum rule.\cite{fcm}
In spite of this shortcoming, the overall shape and temperature dependence of $G$ in these limits
agree with those obtained using NRG.\cite{vel,ander}  

The point that we want to stress here is that although the finite energy features for $T \sim \delta$ and
$\epsilon_F -E_d$ are present not only in $S(T)$ but also in $G(T)$, they are much weaker in $G(T)$. Thus, 
the Seebeck coefficient might be the appropriate tool to study the electronic structure of the system 
at intermediate energies.  

\subsection{Dependence on splitting for $T \rightarrow 0$}                

\label{spli}

\begin{figure}[tbp]
\includegraphics[width=7.5cm]{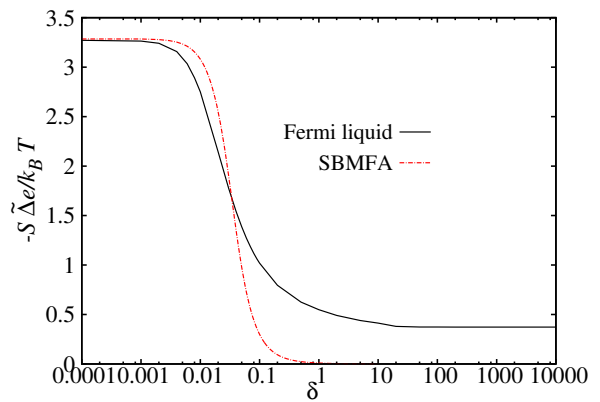} 
\caption{(Color online) Coefficient of the linear dependence of the thermopower as a function of 
the level splitting, for  $E_d=-4 \Gamma$.}
\label{svsd}
\end{figure}
 
In this subsection, we present results for the term linear in temperature $T$
of the thermopower, as $T \rightarrow 0$, using two techniques: SBMFA and 
the Fermi liquid expression Eq. (\ref{sfl}) 
with occupations calculated with the NCA. In Fig. \ref{svsd} we represent $-S/T$ as
a function of $\delta$. In the SU(4) limit $\delta=0$, both techniques agree and 
indicate a very large absolute value of   $S \tilde{\Delta} /T$. In fact, in the extreme Kondo limit of an SU(N) 
model, one has $n_j=1/$N and from Eqs. (\ref{see}), (\ref{in}), (\ref{fl1}), (\ref{fl2}) 
and a Sommerfeld expansion one obtains

\begin{equation}
S = -\frac{\pi^2 T}{3 \tilde{\Delta}}\sin(2\pi/N),  \label{sun}
\end{equation}
and $|S \tilde{\Delta} /T|$ reaches its maximum value $ \pi^2/3=3.29 $ for N=4. The value in Fig. 
\ref{svsd} is slightly smaller due to some degree of intermediate valence.

\begin{figure}[tbp]
\includegraphics[width=7.5cm]{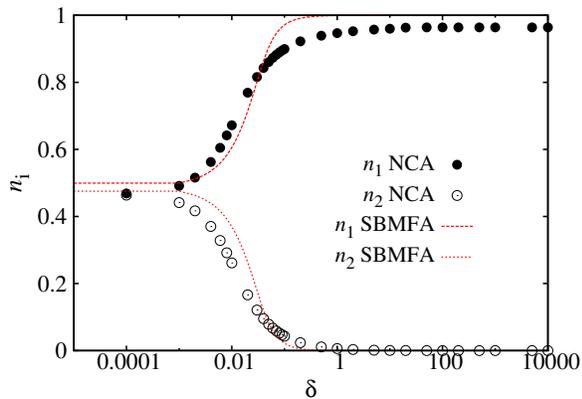} 
\caption{(Color online) Occupation of the two doublets $n_i= \sum_{\sigma} n_{i \sigma}$
as a function of 
the level splitting, for  $E_d=-4 \Gamma$.}
\label{ocup42}
\end{figure}

As $\delta$ increases, there is little variation until the Kondo temperature of the SU(4) limit $T_K^4$ 
($\sim 0.01 \Gamma$ in the figure) is reached. 
For larger $\delta$,  $S \tilde{\Delta} /T$ falls due to the change of occupations
(see Fig. \ref{ocup42}): the lower doublet becomes more populated, 
while the occupation of the higher one decreases,
keeping a total occupation slightly below 1. 
While the trend of the curve is the same for both approaches, NCA and SBMFA, the absolute
value of the thermopower decreases too much within the SBMFA for $\delta > 10 T_K^4$.
This is due to the fact discussed in Section \ref{su2}, that the 
occupation predicted by the SBMFA of the lower lying doublet is too large for $\delta \rightarrow +\infty$.
While this shortcoming affects
the conductance or thermodynamic properties in a few \%, the effect of this increase is more dramatic in the thermopower.

Instead, since the NCA occupations agree with NRG in the limit $\delta \rightarrow +\infty$, 
the approach that combines NCA occupations and Fermi liquid relationships is reliable in this limit.
The agreement with SBMFA and general expectations for the SU(4) model in the Kondo limit, indicates that 
this approach is also reliable for $\delta =0$. 

\section{Summary and discussion}

\label{sum}

We have calculated the thermopower of a model that describes electronic transport through 
quantum dots in C nanotubes,
in which the disorder does not play an essential role, and Si fin-type field effect transistors 
in the Kondo regime. This regime can be easily controlled by applying a gate voltage that 
modifies the energy of the localized levels $E_d$.
Without disorder and applied magnetic field, the model has SU(4) symmetry, as
explained in Section \ref{intro}. In this case, the thermopower as a function of temperature $S(T)$ has a dip
(negative $S$) at temperatures near the Kondo temperature $T_K$, and a peak (positive $S$)
at temperatures near the charge-transfer energy. 
For SU(N) symmetry, varying N with width of the Kondo resonance $\sim 2 T_K$ fixed, 
N=4 is the most favorable case to have a large thermopower
at temperatures lower that the Kondo temperature.

When the SU(4) symmetry is broken, leading to an energy splitting 
$\delta$ between two SU(2) doublets, by a simple
symmetry breaking field (like a magnetic field), a new dip appears in $S(T)$
at temperatures of the order of $\delta$. While this dip the peak 
for positive $S$ at the charge-transfer energy are clearly displayed in $S(T)$, 
the corresponding features in the conductance $G(T)$ at equilibrium are very weak. This 
suggest that the study of the thermopower might be a more useful tool 
to study the electronic structure at finite energies. Another alternative is to
study the conductance out of equilibrium from which peaks at a bias voltage
of the order of $\pm \delta/e$ are expected.\cite{tetta,desint} 

As $\delta$ increases, the characteristic energy scale $T_K(\delta)$, which determines 
among several physical scales, the width of the Kondo resonance and the temperature at
which $G(T)$ falls to half its value at $T=0$, decreases following a simple expression
Eq. (\ref{tk}). For $T \ll T_K(\delta)$, $S$ is linear in $T$ with a coefficient
that stays approximately constant for $\delta < T_K(0)$. As $\delta$ increases further,
$S T_K(\delta)/T$ for $T \rightarrow 0$ decreases by an order of magnitude but 
$T_K(\delta)$ decreases by nearly three orders of magnitude, 
so that the linear coefficient increases. For this calculation, an approach that combines
Fermi liquid relationships with occupation numbers calculated with NCA gives more reliable results
that the SBMFA.

Because of the nature of the underlying SU(4) symmetry, the couplings 
of the four states involved to 
the conducting leads are proportional, and independent of the state
($V_L$ to the left and $V_R$ to the right), This fact which simplifies
the calculations is no more true for weak magnetic field in C nanotubes with 
disorder,\cite{grove} and in general in multilevel quantum dots.\cite{izum2}  
In these cases, as well as in situations with total or partial destructive 
interference, such as transport through molecules,\cite{desint,benzene,mole} the present
formalism does not apply and a full non-equilibrium formalism seems necessary.\cite{benzene} 
In addition here, the total occupation of the dot is below 1, in contrast to models
with multilevel systems in which even occupation is allowed.\cite{serge}

\section*{Acknowledgments}

We thank CONICET from Argentina for financial support. This work was
partially supported by PIP 11220080101821 of CONICET and PICT R1776 of the
ANPCyT, Argentina.
P. R. is sponsored by Escuela de Ciencia y Tecnolog\'{\i}a, Universidad Nacional de San Mart\'{\i}n.


\begin{thebibliography}{99}
\bibitem{tritt} T. M. Tritt, Annu. Rev. Mater. Res. \textbf{41}, 433 (2011).

\bibitem{sales} B. C. Sales, D. Mandrus, and R. K. Williams, 
Science \textbf{272}, 1325 (1996).

\bibitem{tera} I. Terasaki, Y. Sasago, and K. Uchinokura, Phys. Rev. B 
\textbf{56}, R12685 (1997).

\bibitem{bent} A. Bentien, S. Johnsen, G. K. H. Madsen, B. B. Iversen, and
F. Steglich, EPL \textbf{80}, 17008 (2007).

\bibitem{kana} M. G. Kanatzidis, Chem. Mater. \textbf{22}, 648 (2010).

\bibitem{klie} R. F. Klie, Q. Qiao, T. Paulauskas, A. Gulec, A. Rebola, S. 
\"{O}\u{g}\"{u}t, M. P. Prange, J. C. Idrobo, S. T. Pantelides, S. Kolesnik,
B. Dabrowski, M. Ozdemir, C. Boyraz, D. Mazumdar, and A. Gupta, Phys. Rev.
Lett. \textbf{108}, 196601 (2012).

\bibitem{maso} G. D. Mahan, and J. O. Sofo, Proc. Natl. Acad. Sci. USA, 
\textbf{93}, 7436 (1996).

\bibitem{bic} N. E. Bickers, D. L. Cox, and J. W. Wilkins, Phys. Rev. B 
\textbf{36}, 2036 (1987).

\bibitem{velmo} V. Zlati\'{c} and R. Monnier, Phys. Rev. B \textbf{71},
165109 (2005).

\bibitem{cost} T. A. Costi, A. C. Hewson, and V. Zlati\'{c}, J. Phys.:
Condens. Matter \textbf{6}, 2519 (1994); T. A. Costi and A. C. Hewson, 
\textit{ibid.} \textbf{5}, L361 (1993).

\bibitem{vel} T. A. Costi and V. Zlati\'{c}, Phys. Rev. B \textbf{81},
235127 (2010).

\bibitem{harm} T. C. Harman, P. J. Taylor, M. P. Walsh, and B. E. Laforge,
Science \textbf{297}, 2229 (2002).

\bibitem{sche} R. Scheibner, H. Buhmann, D. Reuter, M. N. Kiselev, and L. W.
Molenkamp, Phys. Rev. Lett. \textbf{95}, 176602 (2005).

\bibitem{boese} D. Boese and R. Fazio, Europhys. Lett. \textbf{56}, 576
(2001).

\bibitem{dong} B. Dong and X. L. Lei, J. Phys.: Condens. Matter \textbf{14},
11747 (2002).

\bibitem{kim} T.-S. Kim and S. Hershfield, Phys. Rev. Lett. \textbf{88},
136601 (2002).


\bibitem{hone} J. Hone, I. Ellwood, M. Muno, A. Mizel, M. L. Cohen, A.
Zettl, A. G. Rinzler, and R. E. Smalley, Phys. Rev. Lett. \textbf{80}, 1042
(1998).

\bibitem{yu} C. H.Yu; L. Shi, Z. Yao, D. Y. Li and A. Majumdar, A. Nano
Lett. \textbf{5}, 1842 (2005).

\bibitem{red} P. Reddy, S-Y. Jang, R. A. Segalman, and A. Majumdar, Science 
\textbf{315}, 1568 (2007).

\bibitem{sogo} A. V. Sologubenko, E. Felder, K. Giann\`{o}, H. R. Ott, A.
Vietkine, and A. Revcolevschi, Phys. Rev. B \textbf{62}, R6108 (2000).

\bibitem{sogo2} A. V. Sologubenko, H. R. Ott, G. Dhalenne, and A.
Revcolevschi, Europhys. Lett. \textbf{62}, 540 (2003).

\bibitem{roz} A. V. Rozhkov and A. L. Chernyshev, Phys. Rev. Lett. \textbf{94},
087201 (2005).

\bibitem{arra} L. Arrachea, G. S. Lozano, and A. A. Aligia, Phys. Rev. B 
\textbf{80}, 014425 (2009).

\bibitem{lim} J. S. Lim, M.-S. Choi, M.\thinspace Y. Choi, R. L\'{o}pez, and
R. Aguado, Phys. Rev. B \textbf{74}, 205119 (2006).

\bibitem{ander} F. B. Anders, D. E. Logan, M. R. Galpin, and G. Finkelstein,
Phys. Rev. Lett. \textbf{100}, 086809 (2008).

\bibitem{lipi} S. Lipinski and D. Krychowski, Phys. Rev. B \textbf{81},
115327 (2010).

\bibitem{buss} C. A. B\"{u}sser, E. Vernek, P. Orellana, G. A. Lara, E. H.
Kim, A. E. Feiguin, E. V. Anda, and G. B. Martins, Phys. Rev. B \textbf{83},
125404 (2011).

\bibitem{fcm} L. Tosi, P. Roura-Bas, and A. A. Aligia, 
Physica B \textbf{407}, 3259 (2012).

\bibitem{grove} K. Grove-Rasmussen, S. Grap, J. Paaske, K. Flensberg, S.
Andergassen, V. Meden, H. I. Jorgensen, K. Muraki, and T. Fujisawa,
Phys. Rev. Lett. \textbf{108}, 176802 (2012).

\bibitem{tetta} G. C. Tettamanzi, J. Verduijn, G. P. Lansbergen, M.
Blaauboer, M. J. Calder\'{o}n, R. Aguado, and S. Rogge, Phys. Rev. Lett. 
\textbf{108}, 046803 (2012).

\bibitem{he} Y. He and G. Galli, Phys. Rev. Lett. \textbf{108}, 215901
(2012).

\bibitem{col} P. Coleman, Phys. Rev. B \textbf{29}, 3035 (1984).

\bibitem{hew} A. C. Hewson, \textit{The Kondo Problem to Heavy Fermions, }%
Cambridge University Press, Cambridge, England, 1993 .

\bibitem{lady} A. A. Aligia and L. A. Salguero, Phys. Rev. B \textbf{70},
075307 (2004); Phys. Rev. B \textbf{71}, 169903(E).

\bibitem{soc} A. M. Lobos and A. A. Aligia, Phys. Rev. Lett. \textbf{100},
016803 (2008); Physica B \textbf{404}, 3306 (2009).

\bibitem{compa} T. A. Costi, J. Kroha and P. W\"{o}lfle, Phys. Rev. B 
\textbf{53}, 1850 (1996).

\bibitem{roura} P. Roura-Bas, Phys. Rev. B \textbf{81}, 155327 (2010).

\bibitem{ogu} A. Oguri, J. Phys. Soc. Jpn. \textbf{74}, 110 (2005).

\bibitem{rpt} J. Rinc\'{o}n, A. A. Aligia, and K. Hallberg, Phys. Rev. B 
\textbf{79}, 121301(R) (2009); Phys. Rev. B \textbf{80}, 079902(E) (2009);
Phys. Rev. B \textbf{81}, 039901(E) (2010).

\bibitem{sela} E. Sela and J. Malecki, Phys. Rev. B \textbf{80}, 233103
(2009).

\bibitem{scali} A. A. Aligia, J. Phys. Condens. Matter \textbf{24}, 015306
(2012); references therein.

\bibitem{bulla} R. Bulla, A.C. Hewson, and Th. Pruschke, 
J. Phys. Cond. Matt.  \textbf{10}, 8365 (1998).

\bibitem{benzene} L. Tosi, P. Roura-Bas, and A. A. Aligia, 
J. Phys. Condens. Matter \textbf{24}, 365301 (2012).

\bibitem{desint} P. Roura-Bas, L. Tosi, A. A. Aligia, and K. Hallberg, Phys.
Rev. B \textbf{84}, 073406 (2011).

\bibitem{baru} P. P. Baruselli, A. Smogunov, M. Fabrizio, and E. Tosatti,
Phys. Rev. Lett. \textbf{108}, 206807 (2012).

\bibitem{meir} Y. Meir and N. S. Wingreen, Phys. Rev. Lett. \textbf{68},
2512 (1992).

\bibitem{yoshi2} A Yoshimori and A Zawadowski, J. Phys. C \textbf{15}, 5241 (1982).

\bibitem{yoshi} A. Yoshimori, Prog. Theor. Phys. \textbf{55}, 67 (1976).

\bibitem{kirchner} S. Kirchner, J. Kroha, and P. W\"{o}lfle, Phys. Rev. B 
\textbf{70}, 165102 (2004).

\bibitem{logan} D. E. Logan, M. P. Eastwood and M. A. Tusch, J. Phys.:
Condens. Matter \textbf{10}, 2673 (1998).

\bibitem{vau} L. Vaugier, A.A. Aligia and A.M. Lobos, Phys. Rev. B \textbf{76}, 165112 (2007). 

\bibitem{roch} N. Roch, S. Florens, V. Bouchiat, W. Wernsdorfer, and F.
Balestro, Nature \textbf{453}, 633 (2008).

\bibitem{serge} S. Florens, A, Freyn, N. Roch, W. Wernsdorfer, F. Balestro,
P. Roura-Bas and A. A. Aligia, J. Phys. Condens. Matter \textbf{23}, 243202
(2011); references therein.

\bibitem{st1} P. Roura-Bas and A. A. Aligia, Phys. Rev. B \textbf{80},
035308 (2009).

\bibitem{st2} P. Roura-Bas and A. A. Aligia, J. Phys. Cond. Matt. \textbf{22}, 025602 (2010).

\bibitem{freyn} A. Freyn and S. Florens, Phys. Rev. Lett. \textbf{107},
017201 (2011).

\bibitem{izum2} W. Izumida, O. Sakai, and Y. Shimizu, J. Phys. Soc. Jpn. 
\textbf{67}, 2444 (1998).

\bibitem{mole} J. Rinc\'{o}n, K. Hallberg, A. A. Aligia, and S. Ramasesha,
Phys. Rev. Lett. \textbf{103}, 266807 (2009); references therin.



\end{thebibliography}
\end{document}